\documentclass[prl,twocolumn,showpacs, amssymb, amsmath,aps] {revtex4}

\usepackage{graphicx}
\usepackage{dcolumn}
\usepackage{bm}
\usepackage{epstopdf}
\usepackage{latexsym}
\usepackage{mathrsfs}
\begin{document}


\title{Quantum Control of the Hyperfine Spin of a Cs Atom Ensemble}

\author{Souma Chaudhury, Seth Merkel, Tobias Herr, Andrew Silberfarb, Ivan H. Deutsch, and Poul S. Jessen}
\affiliation{%
College of Optical Sciences, University of Arizona, Tucson, Arizona 85721, USA\\
Department of Physics and Astronomy, University of New Mexcico, Albuquerque, New Mexico 87131, USA
}%

\date{\today}

\begin{abstract}
We demonstrate quantum control of a large spin-angular momentum associated with the $F=3$ hyperfine ground state of $^{133}$Cs.  A combination of time dependent magnetic fields and a static tensor light shift is used to implement near-optimal controls and map a fiducial state to a broad range of target states, with yields in the range 0.8-0.9.  Squeezed states are produced also by an adiabatic scheme that is more robust against errors.  Universal control facilitates the encoding and manipulation of qubits and qudits in atomic ground states, and may lead to improvement of some precision measurements. \end{abstract}


\maketitle

Accurate dynamical control plays a central role when quantum mechanics is leveraged to improve the outcome of some physical process. Quantum control has been accomplished in many contexts and at various levels of sophistication, and has become a mainstay in areas such as nuclear magnetic resonance \cite{chuangrevmod}, coherent chemistry \cite{shapiro}, quantum information processing \cite{nielsenchuang} and quantum metrology \cite{maccone}.  One extensively studied problem is how to transfer a physical system from an initial fiducial state (e.g. the ground state) to some final state, as is done, for example, in optical control of chemical reactions \cite{shapiro}. In such cases the figure of merit for control is the yield, or overlap between the actual and desired states. As long as errors and decoherence are negligible the general topography of control landscapes (yield vs control parameters) is well understood \cite{rosenthal} and techniques are available for efficient design of optimal controls \cite{rabitz2003}. The most ambitious level of quantum control requires that the system be \emph{controllable} in the Lie-algebraic sense \cite{ramakrishna}, a sufficient condition for which is that internal dynamics plus interaction with external fields can generate any unitary transformation within the Hilbert space of interest. Even when full control is possible in principle, attention must be paid to robustness in the presence of realistic levels of dissipation and systematic errors in the control fields. In spin-1/2 systems this can be accomplished through open loop control \cite{khaneja}, i. e. without recourse to real-time feedback \cite{doherty2000} or error correction \cite{nielsenchuang}, but little is known about how to systematically design robust controls in a larger state space.

In this letter we demonstrate quantum control of the spin-angular momentum (nuclear plus electronic) associated with the $F=3$  hyperfine ground state of individual $^{133}Cs$ atoms, corresponding to a $2F+1=7$ dimensional Hilbert space. Starting from an easily prepared fiducial state we use magnetic fields and AC Stark shifts (light shifts) to design and implement near-optimal controls and produce a range of target states. We evaluate our control performance by experimentally reconstructing the entire spin density matrix \cite{silberfarb} and computing the overlap between the measured and target states. In most cases the estimated yield is in the $0.8-0.9$ range, limited by errors in the control fields and to a lesser extent by decoherence from light scattering.  The measured states can be compared also to the predictions of a full model that includes the effects of errors and decoherence. Typical fidelities between measured and predicted states are around $0.9$, which is close to the resolution limit of our procedure for quantum state estimation. We further use this universal approach to generate spin-squeezed states and compare against a method based on adiabatic evolution \cite{molmersorensen}. The latter is more robust against errors in the control fields, but also slower and thus more sensitive to light scattering and decoherence. Large spins provide a testing ground for the design of accurate and robust controls in a system where the Hamiltonian is well known and where errors and dissipation are well understood and can be accurately modeled. From a practical perspective, quantum control of hyperfine states has direct relevance to proposals for neutral atom quantum computing \cite{jessenQIP2004} wherein qubits (or higher dimensional qudits \cite{bullock2005}) are encoded in the ground-state manifold, and may provide a simple route to modest spin squeezing and accompanying gains in precision atomic magnetometry \cite{geremia2005}.

\begin{figure*}
[t]\resizebox{17.5cm}{!}
{\includegraphics{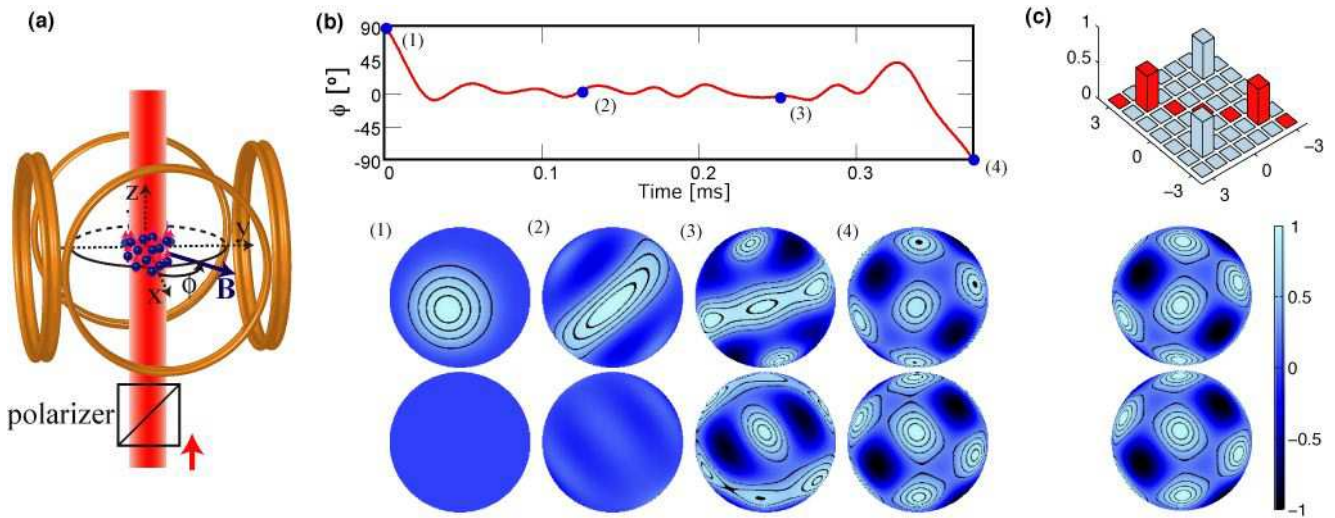}}
\caption{\label{fig:setup} Quantum control of a large atomic spin.  (a) Schematic of the experiment.  (b) Example of a control waveform $\phi(t)$. (1-4) Wigner functions at four stages during the control sequence. Both sides of the sphere are shown.  The final result is close to the target state $ \vert \chi_T \rangle = (\vert m_z=2 \rangle + \vert m_z=-2 \rangle)/\sqrt{2}$. (c) Density matrix (absolute values) and  Winger function for $ \vert \chi_T \rangle$}
\end{figure*}

Universal control of a spin $F$ requires that the Hamiltonian dynamics be capable of generating an arbitrary unitary transformation in $SU(2F+1)$. A linear Zeeman interaction between the atomic magnetic moment and a moderate strength magnetic field generates only Larmor precession and the geometric rotations that are a representation of $SU(2)$. More general control requires a Hamiltonian that is non-linear in at least one component of $\mathbf{F}$. In our experiment this is provided by an off-resonance light field that couples to the atomic ground state through the tensor ac-polarizability and leads to a spin-dependent light shift with an irreducible rank-2 component \cite{geremia}. The combination of a time-dependent magnetic field and a constant x-polarized light field results in a control Hamiltonian \cite{smithnonlinear},

\begin{equation}
\hat{H}_C(t) = 
g_{f}\mu_{B} \mathbf{B}(t) \mathbf{\cdot \hat{F}} +
\beta \hbar \gamma_s F_x^2
\label{eq:controlH}
\end{equation}

where we have expressed the strength of the nonlinearity in terms of the photon scattering rate $\gamma_{s}$ and where the dimensionless parameter $\beta$ is a measure of the timescales for coherent versus incoherent evolution. Its value depends on the atomic structure and the frequency of the driving field and for Cs takes on a maximum value $\beta=8.2$ when tuned between the hyperfine transitions of the $D_1$  line at $894 nm$. This is enough to allow considerable coherent manipulation. It follows directly from the theory of Lie groups that a Hamiltonian of this form allows full control of a spin of any magnitude \cite{lloyd}. Specifically, one can show that the algebra generated by commutators and linear combinations of  ${F_x,F_y,F_x^2}$ spans the entire $(2F+1)^2 - 1$ dimensional operator space necessary to represent $SU(2F+1)$. Thus, a time-varying magnetic field in the $x-y$ plane suffices to make $\hat{H}_C(t)$ universal.

A schematic of our setup for spin quantum control is shown in Fig.~\ref{fig:setup}(a). We begin with a sample of a few million Cs atoms, captured and laser cooled to $\sim2 \mu K$ in a magneto-optical trap and optical molasses. Once the atoms are released from the optical molasses their spin state is initialized by optical pumping into a state of maximum projection along the $y$-axis, $|\psi_0\rangle = |F=3,m_y=3\rangle$. We drive the spins by applying a time-dependent magnetic field from a set of low-inductance coils driven by arbitrary waveform generators, and by applying a static light shift from an optical probe beam. Using an all-glass vacuum cell, avoiding nearby conductive or magnetizable materials, and synchronizing our $\sim0.5 ms$ duration experiment to a fixed point during the AC line cycle allows us to null the background magnetic field to a few tens of $\mu$Gauss without the use of shielding or active compensation. The applied magnetic field can be controlled with an accuracy better than one percent in a bandwidth of more than $100 kHz$. Immediately following a period of quantum control we estimate the resulting quantum state as described in \cite{silberfarb}. In this procedure the control magnetic and optical fields are applied to drive the spins for an additional 1.5 ms, while continually and weakly measuring a spin observable through its coupling to the probe polarization. To reduce the effect of noise, the measurement signal is averaged over 16 repetitions of the experiment and the full density matrix determined from the measurement record and the known evolution.

Control Hamiltonians for our experiment are designed through a simple procedure that we have found empirically to produce Òvery goodÓ though not provably optimal results. The objective is to start from the state $|\psi_0\rangle$ and to produce a specified target state $|\chi_T\rangle$ by modulating the field $\mathbf{B}(t)$ for a fixed time $\tau$. With readily available magnetic fields the timescale for geometric rotations is much shorter than for nonlinear evolution driven by the light shift, and the latter therefore becomes the time-limiting element of most transformations. In our experiment the maximum available Larmor frequency is $15 kHz$ and the nonlinear strength is $\beta \gamma_S \approx 2\pi \times 500Hz$. Under these conditions there is no significant sacrifice in control performance when the set of available rotations is somewhat restricted. We therefore choose the magnetic field to have constant magnitude and time-dependent direction in the $x$-$y$ plane. With this simplification the control Hamiltonian is completely determined by the time dependent angle $\phi (t)$ between $\mathbf{B}(t)$  and the $x$-axis. The state $|\chi_T\rangle$ and the transformation $|\psi_0\rangle \to |\chi_T\rangle$ belong to a $d=7$ dimensional Hilbert space and can be specified by a set of $2d-2=12$ real numbers, and full control therefore requires at least that many free parameters in the control Hamiltonian. To ensure sufficient flexibility we specify the control waveform $\phi(t)$ by its values $\{ \phi_i\}$ at $N=30$ discrete time steps.

\begin{figure}
[t]\resizebox{8.75cm}{!}
{\includegraphics{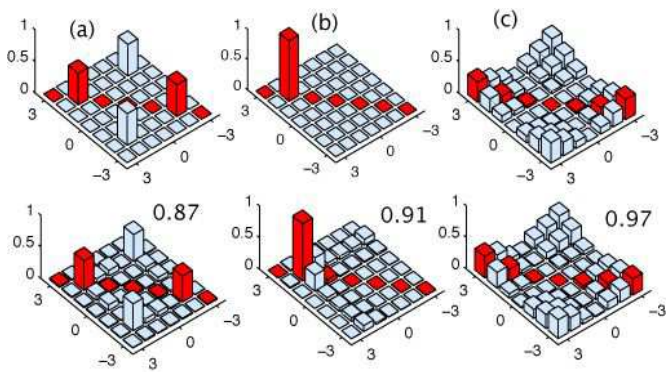}}
\caption{\label{fig:estimates}Examples of target and measured density matrices (absolute values). The target states are (a) $(\vert m_z=2 \rangle + \vert m_z=-2 \rangle)/\sqrt{2}$ , (b)$\vert m_x=2\rangle$, (c) $\mathbf{\Sigma}_{y}m_y\vert m_y\rangle$. The experimental yield is indicated for each case.}
\end{figure}

The design of a control waveform now proceeds through two search iterations. In the first round we calculate the state $|\psi_P\rangle$ produced by a sequence of field directions $\{ \phi_i\}$ by integrating the Schršdinger equation for a time $\tau$, with suitable filtering of the corresponding $\mathbf{B}(t)$ to reflect the bandwidth and slew rate limitations of our magnetic coils and drivers. A locally optimal control waveform is found by starting from a random seed and maximizing the yield $\mathscr{Y}=|\langle \chi_T| \psi_P \rangle|^2$ with a simple gradient ascent algorithm. We have found that a small set of random seeds almost always generates at least one waveform with yield greater than 0.99, which is expected from the general structure of control landscapes derived in \cite{rosenthal}. At this point we switch to a more realistic estimate of control performance by modeling the evolution with a full master equation that incorporates decoherence from light scattering and inhomogeneity of the nonlinear strength across the atomic ensemble. This allows a second stage of optimization starting from the waveform generated in round one and using the more complete but computationally intensive model to predict the yield, which is now defined in terms of the overlap $\mathscr{Y} = Tr\sqrt {\rho_T^{1/2} \rho_{P}^{} \rho_T^{1/2}}$ between the target density matrix $\rho_T^{}$ and the predicted density matrix $\rho_P^{}$.

An example of an optimized control waveform is shown in Fig.~\ref{fig:setup}(b), along with Wigner function representations of the spin ÒwavepacketÓ \cite{agarwal} at a few steps during the transformation as calculated using the complete master equation. Note that the nonlinear evolution initially produces a squeezing ellipsoid which later wraps around the sphere so that interference effects can be manipulated to create the desired state. The end product is very close to the target state shown in Fig.~\ref{fig:setup}(c). According to our model this and a wide variety of other control waveforms all produce yields near 0.95. Taking into account imperfect optical pumping in our experiment (the initial population in $|\psi_0\rangle$ is $\sim0.96$) reduces the expected yields to around 0.90.

\begin{figure}
[t]\resizebox{8.75cm}{!}
{\includegraphics{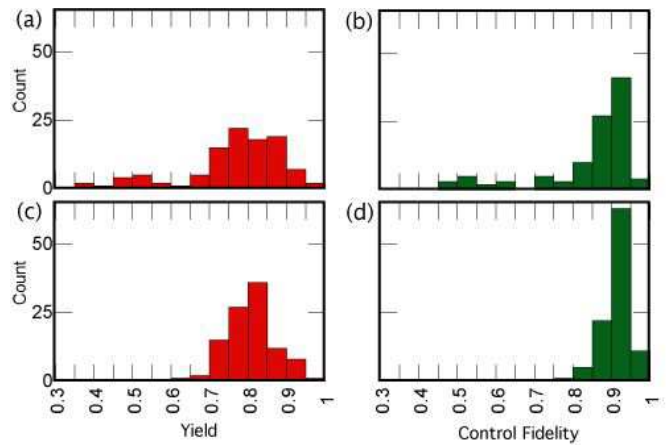}}
\caption{\label{fig:histogram} Histograms of (a) yields and (b) fidelities of measured vs. predicted states. (c) \& (d) Yields and fidelities when each measured state is geometrically rotated to optimize overlap with the predicted state.}
\end{figure}

We have generated and tested a sample of control waveforms designed to produce 21 different pure spin states. Fig.~\ref{fig:estimates} shows three examples of target and measured density matrices, with yields falling in the range 0.87-0.97. A more complete statistics of yields for over a hundred experimental realizations of control is compiled in the form of a histogram in Fig.~\ref{fig:histogram}(a), showing a fairly broad distribution centered on respectable value of 0.8. It is also informative to compare the measured density matrices $\rho_M^{}$ against the density matrices $\rho_P^{}$ predicted by our model, as quantified by the fidelity $\mathcal{F} = Tr\sqrt {\rho_P^{1/2} \rho_M^{} \rho_P^{1/2}}$. Fig.~\ref{fig:histogram}(b) shows a histogram of fidelities for our data set. Note that both yield and fidelity can be affected by control errors (the real state is different from $\rho_P^{}$) as well as state estimation errors (the real state is different from $\rho_M^{}$), and that there is no way to distinguish between these possibilities. Numerical modeling shows that small background magnetic fields or miscalibration of the control fields will lead to apparent geometric rotations of the final state, but such errors are too small in our experiment to significantly affect the outcome. The obvious outliers in the yield and fidelity distributions are associated with two specific control waveforms, and closer examination shows that the estimated states are rotated relative to the predicted states. The axis of rotation corresponds the direction of the magnetic field at the transition between the control and state estimation phases, which suggests a problem with the way the corresponding control waveforms were joined together. We can numerically rotate a given $\rho_M^{}$ to maximize its fidelity relative to $\rho_P^{}$ and obtain new values for yield and fidelity. Carrying out this procedure for all data points takes care of the outliers without otherwise changing the yield distribution significantly, as shown in Fig.~\ref{fig:histogram}(c). This distribution can reasonably be interpreted as a measure of our ability to control the spins in a well designed experiment. The fidelity distribution (Fig.~\ref{fig:histogram}(d)) remains peaked at ~0.9, which we know from experience to reflect the accuracy of our state estimation algorithm. Finally we note that random errors in state estimation are far more likely to decrease than increase the apparent yield. A simple error model based on Gaussian random displacements in state space indicate that the yields are probably $10\%$ larger on average than indicated by Fig.~\ref{fig:histogram}(d). This puts most yields in the range 0.8-0.9, in good agreement with the $\sim 0.9$ predicted by the model used to design the control waveforms in the first place.

\begin{figure}
[t]\resizebox{8.75cm}{!}
{\includegraphics{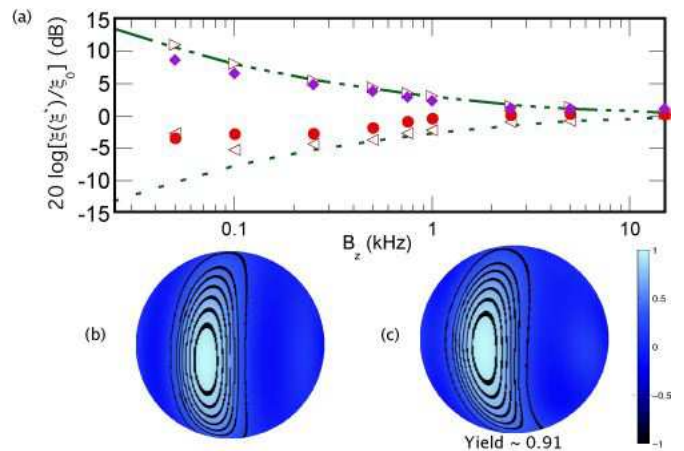}}
\caption{\label{fig:squeezing} (Color online) Spin squeezing by adiabatic control. (a)  Normalized squeezing parameter vs. final magnetic field for the squeezed and anti-squeezed components. Dashed lines: perfect squeezing. Open symbols: predictions from our theoretical model. Filled symbols: experimental results. (b) Target and (c) measured Wigner functions corresponding to the smallest observed $\xi$.}
\end{figure}

To further explore quantum control in our system we have studied the generation of spin squeezing both by optimal control as outlined above and by the adiabatic scheme described in \cite{molmersorensen}. The latter begins with an initial state, $|\psi_0\rangle = |F=3,m_y=-3\rangle$, which has equal uncertainties for the components $\Delta F_x$ and $\Delta F_z$ and is often referred to as a spin-coherent state. This state is a good approximation to the ground state of the control Hamiltonian $\hat{H}_C(t)$ when the magnetic field is of the form $\mathbf{B}(t)=B(t)\mathbf{y}$ and $B(t)$ is large. As the field magnitude is slowly reduced the state adiabatically evolves so as to minimize the squeezing parameter $\xi = \Delta F_x/|\langle F_y \rangle|$ of relevance for metrology \cite{wineland}. Fig.~\ref{fig:squeezing}(a) shows the progression of squeezing and anti-squeezing relative to a spin-coherent state with the same $|\langle F_y \rangle|$. Up to $\sim 4 dB$ of squeezing is seen in the experiment, in good agreement with the predictions of our model. For the small spin magnitude used here the squeezing is quickly limited by the decrease in $|\langle F_y \rangle|$ as the squeezing ellipse wraps around the sphere. Fig.~\ref{fig:squeezing}(b)-(c) shows Wigner functions for the target and actual state for the smallest $\xi$ achieved in our experiment ($\sim 80\%$ of the coherent state value). We have produced the same spin squeezed states via optimal control, with small but significant reductions in both squeezing and yield. This suggests that gains from reduced decoherence (optimal control is as much as five times faster) is offset by increased sensitivity to control errors.

In conclusion we have implemented a scheme for optimal control of the hyperfine spin-angular momentum of Cs atoms in the $F=3$ ground state. Control Hamiltonians were designed to produce a range of target states, applied in the laboratory and evaluated by measuring the resulting density matrices. Typical yields fall in the range 0.8-0.9. Among the states produced were a number of spin squeezed states, which allowed  direct comparison of our optimal control approach to an adiabatic scheme that is more robust to errors in the control fields. In future experiments we plan to use a combination of rf and microwave magnetic fields to control the entire 16 dimensional state space for the Cs $6S_{1/2}(F=3,4)$ ground manifold. Preliminary studies indicate that this system is fully controllable on a timescale of a few tens of microseconds with easily available control fields. This will provide an important tool for the encoding and manipulation of qubits and qudits embedded in a larger atomic ground manifold. In the longer term it is also interesting to consider if control Hamiltonians of the form used here can be achieved for collective spins, for example through coherent optical feedback \cite{takeuchi} or through atom-atom interactions in a quantum-degenerate gas \cite{micheli2003}.

This research was supported by grants NSF Nos. PHY-0355073 and PHY-0355040, ONR No. N00014-05-1-420 and DTO No. DAAD19-13-R-0011.


\end{document}